\documentclass[preprint2]{aastex61}

\usepackage{graphicx}
\usepackage{amsmath}
\usepackage{natbib}
\usepackage{mathtools}
\usepackage{color}
\usepackage{txfonts}

\newcommand{\sdoaia}{{\it SDO}/AIA\ }

\newcommand{\muhz}{$\mu$Hz}

\submitjournal{ApJ}

\shorttitle{The Coronal Monsoon}
\shortauthors{Auch\`ere et al.}

\begin{document}

\title{The Coronal Monsoon: Thermal Nonequilibrium Revealed by Periodic Coronal Rain}
\correspondingauthor{Fr\'ed\'eric Auch\`ere}
\email{frederic.auchere@ias.u-psud.fr}
\newcommand{\iasaffil}{Institut d'Astrophysique Spatiale, CNRS, Univ. Paris-Sud, Universit\'e Paris-Saclay, B\^at. 121, F-91405 Orsay, France}
\newcommand{\iacaffil}{Institute of Applied Computing \& Community Code, Universitat de les Illes Balears, E-07122 Palma de Mallorca, Spain}

\author[0000-0003-0972-7022]{Fr\'ed\'eric Auch\`ere}
\affiliation{\iasaffil}
\author[0000-0001-5315-2890]{Clara Froment}
\affiliation{Rosseland Centre for Solar Physics, University of Oslo, P.O. Box 1029 Blindern, NO-0315 Oslo, Norway}
\affiliation{Institute of Theoretical Astrophysics, University of Oslo, P.O Box 1029, Blindern, NO-0315, Oslo, Norway}
\author[0000-0001-9295-1863]{Elie Soubri\'e}
\affiliation{\iasaffil}
\affiliation{\iacaffil}
\author[0000-0003-1529-4681]{Patrick Antolin}
\affiliation{School of Mathematics and Statistics, University of St. Andrews, St. Andrews, Fife KY16 9SS, UK}
\author[0000-0003-4162-7240]{Ramon Oliver}
\affiliation{\iacaffil}
\affiliation{Departament de F\'isica, Universitat de les Illes Balears, E-07122 Palma de Mallorca, Spain}
\author[0000-0002-0397-2214]{Gabriel Pelouze}
\affiliation{\iasaffil}
\date{Received 2017 November 20; revised 2017 December 18; accepted 2018 January 3}

\begin{abstract}
We report on the discovery of periodic coronal rain in an off-limb sequence of {\it Solar Dynamics Observatory}/Atmospheric Imaging Assembly images. The showers are co-spatial and in phase with periodic (6.6~hr) intensity pulsations of coronal loops of the sort described by \cite{Auchere2014} and \cite{Froment2015, Froment2017}. These new observations make possible a unified description of both phenomena. Coronal rain and periodic intensity pulsations of loops are two manifestations of the same physical process: evaporation / condensation cycles resulting from a state of thermal nonequilibrium (TNE). The fluctuations around coronal temperatures produce the intensity pulsations of loops, and rain falls along their legs if thermal runaway cools the periodic condensations down and below transition-region (TR) temperatures. This scenario is in line with the predictions of numerical models of quasi-steadily and footpoint heated loops. The presence of coronal rain -- albeit non-periodic -- in several other structures within the studied field of view implies that this type of heating is at play on a large scale.
\end{abstract}
\textbf{}

\section{Context\label{sec:context}}

\begin{figure*}[t!]
\includegraphics[width=\textwidth]{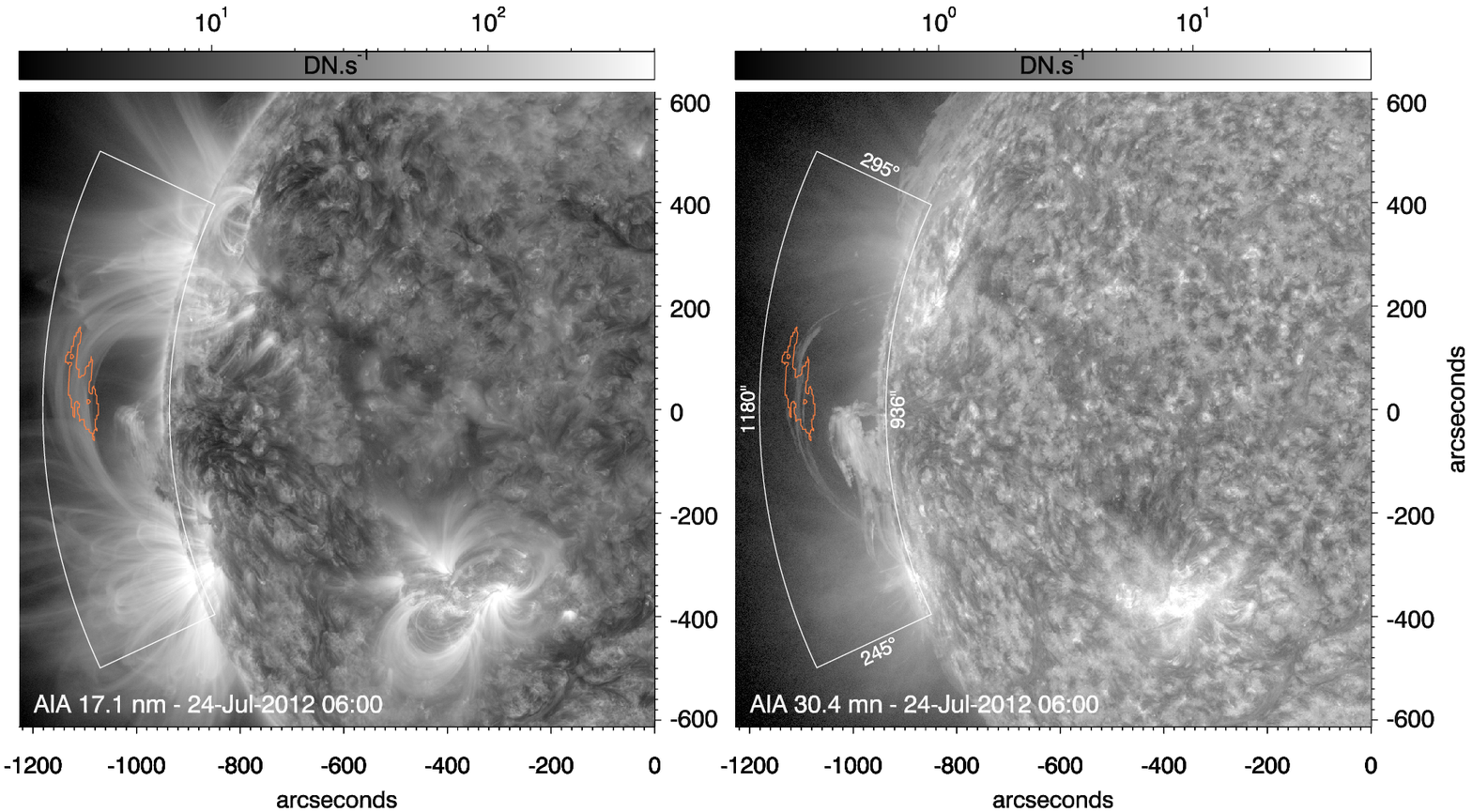}
\caption{The ROI overplotted in white on two simultaneous 17.1~nm (left) and 30.4~nm (right) AIA context images taken at the middle of the sequence. Excess Fourier power is detected inside the orange contour (Figure \ref{fig:power_map}).\label{fig:fov}}
\end{figure*}

Discovered in the 1970s \citep{Kawaguchi1970, Leroy1972}, coronal rain is defined observationally as transient, elongated, fine-structured blob-like features falling along coronal loops and visible mostly off-limb in cool (chromospheric) spectral lines \citep{Schrijver2001, DeGroof2004, DeGroof2005, Antolin2010, Antolin2011, Antolin2012, Vashalomidze2015}. These blobs are now understood from many numerical simulations -- either hydrodynamic in one dimension \citep[1D,][]{Kuin1982, Martens1983, Antiochos1991, Karpen2001, Karpen2005, Muller2003, Muller2004, Karpen2008, Xia2011, Mikic2013} or magnetohydrodynamic (MHD) in 1.5 \citep{Antolin2010}, 2.5 \citep{Fang2013, Fang2015} or 3-dimensions \citep[3D,][]{Moschou2015, Xia2017} -- to be cold condensations formed by runaway cooling: in the corona, the increasing radiative losses with decreasing temperature causes a positive feedback if the heat input is insufficient to maintain the energy balance. This  happens in coronal loops heated steadily (which encompasses the quasi-steady case of impulsive heating with interruptions shorter than the cooling time) and predominantly at their footpoints. Under these conditions a loop is in a state of thermal nonequilibrium (TNE) in which no equilibrium exists \citep{Antiochos1999}, resulting in an intrinsically dynamic behavior. Heating concentrated at both ends of a low-density flux tube progressively fills it with evaporated hot plasma. The mass of plasma builds up so that at some point the thermal conduction does not transport enough heat along the loop to sustain its temperature, which locally renders the plasma thermally unstable \citep{Parker1953, Field1965}. Any additional mass then initiates runaway cooling and the ensuing condensation - the ``rain'' - rapidly grows to ultimately fall into the chromosphere, thus draining the loop down to its initial lower density. In the flare-driven scenario, the rain is triggered in otherwise quiescent loops by a single and long enough burst of intense footpoint heating \citep{Scullion2016}. But if both the geometry and the heating are steady, the process repeats itself over and over again.

This periodic behavior is a prominent feature of most of the above-cited 1D numerical simulations of TNE because they impose a static geometry and an ad-hoc stratified and constant heating function. Similarly, all multi-dimensional simulations to date impose static or quasi-static magnetic boundary conditions and a heating that is usually a function of the magnetic field strength. These setups produce long-lasting magnetic loop bundles (with little geometric evolution) that develop TNE cycles if the model is run for a long enough time, as in \citet{Fang2015}. However, while the periodicity of the plasma response is a strong prediction of the models (1D-3D), the existence in reality of persistent TNE cycles and of the associated periodic coronal rain in the same coronal loop bundle is not straightforward. Both observations and self-consistent 3D MHD large-scale simulations indicate, respectively, strong variability of the intensity \citep{Reale2014} and connectivity of magnetic field lines \citep{Gudiksen2005b} that seemingly make it unlikely that TNE could be maintained for several cycles. To the best of our knowledge, while coronal rain is a recurring phenomenon \citep{Antolin2012}, it has never been reported to be periodic. This is either because the observations are usually significantly shorter than the duration of a TNE cycle (several hours or tens of hours), or because the natural variations of the heating and geometry produce cycles of varying duration and possibly the return to equilibrium.
 
Processing more than 13 years of quasi-continuous 12 minute-cadence observations with the 19.5~nm passband of the {\it Solar and Heliospheric Observatory} \citep{Domingo1995} Extreme-ultraviolet Imaging Telescope \citep{Delaboudiniere1995}, we discovered 499 periodic (3-16 hr) intensity pulsation events in on-disk active regions, 268 of which are clearly associated with loops, and some lasting for up to six days \citep{Auchere2014}. \cite{Froment2016} found more than 2000 similar events in six years of images taken in the six coronal passbands of the {\it Solar Dynamics Observatory} \citep[{\it SDO};][]{Pesnell2012} Atmospheric Imaging Assembly \citep[AIA;][]{Lemen2012}. For three of these events, \cite{Froment2015} demonstrated by differential emission measure (DEM) analysis \citep{Guennou2012a, Guennou2012b, Guennou2013} that the intensity pulsations were caused by periodic fluctuations of the temperature and emission measure (EM), with the EM peaking after the temperature, as expected in the above-described scenario. \cite{Froment2017} went on to numerically simulate the TNE state of one of these loops and successfully reproduced the observed light curves. The loop geometry and heating conditions can thus be steady enough to maintain a state of TNE for at least 15 of these 9~hr  cycles. However, while it emerges that coronal rain and intensity pulsations are two different manifestations of the same underlying process, coronal rain could not be detected as might have been expected in the 30.4~nm passband of \sdoaia during the coolest phases of the cycles. This is either because the loops were in a specific regime of TNE in which the condensations are siphoned away before they can reach lower TR temperatures \citep[as suggested by][]{Froment2017}, 
or because of detection limitations, like rain occurring at spatial scales too small to be resolved by AIA, or poor distinguishability of the blobs against the fine-structured underlying background.

In this paper, we present new observations of coronal loops in thermal nonequilibrium that exhibit both intensity pulsations and periodically recurring coronal rain, thus definitely unifying the two phenomena. The selected event was dubbed the ``Rain Bow'' because the ``rain'' occurs along a large ``bow'' of  loops. It is described in Section~\ref{sec:rainbow}, which is organized as follows:
\begin{enumerate}
\item \S~\ref{sec:signal_properties} describes the properties of the periodic signal:
\subitem \S\ref{sec:detection} details the statistics used to establish the significance of the detection.
\subitem \S\ref{sec:pulses} shows that the observed periodic signal is not the signature of a vibration mode but that of a cyclical process.
\subitem \S\ref{sec:haze} explains the effect of background and foreground emission on the signal strength in the different AIA passbands.
\item \S~\ref{sec:cooling} demonstrates that the loops are cooling periodically. At this point, we conclude that the present coronal pulsations are an off-disk equivalent to those observed on-disk by \citet{Froment2015}.
\item \S\ref{sec:rain} then describes the 30.4~nm rainfalls occurring during the coolest parts of the TNE cycles.
\end{enumerate}

Finally, our findings and their interpretation are summarized in \S~\ref{sec:conclusions}.

\section{The ``Rain Bow''\label{sec:rainbow}}

The event described below was selected among those discovered\footnote{About 3000 events have been detected. Statistics similar to those given in \cite{Auchere2014} and \cite{Froment2016} for the on-disk events are being worked out.} in the archive of AIA images using the automated detection code developed by \cite{Auchere2014}, modified to run off-disk since coronal rain is otherwise notably difficult to observe with imaging instruments alone \citep{Antolin2012, Antolin2012b}. The feature tracking in heliographic coordinates was thus switched off and the temporal intervals limited to 2.5 days in order to limit the apparent height variation of the structures as they rotate around the limb. For a given region of interest (ROI), the code triggers on the presence of clusters of significant coherent power in the corresponding cube of Fourier power spectra.

\subsection{Properties of the periodic signal\label{sec:signal_properties}}

In this section and the next, we redo the detection analysis as in the automated code but at higher spatial and temporal resolutions, and we demonstrate that the observed off-disk coronal pulsations have spectral properties identical to those of the on-disk events of \cite{Froment2015, Froment2017}. The 30.4~nm signal associated with the rain will be discussed in \S~\ref{sec:rain}.

The ``Rain Bow'' event was initially detected at 17.1~nm in the period from 2012 July 23 at 00:00 UT to 2012 July 25 at 12:00 UT. For the following analysis, the ROI is restricted to the sector of Sun-centered annulus enclosing the system of large trans-equatorial loops visible above the east limb (Figure \ref{fig:fov}). The ROI is sampled on a regular $1250 \times 400$ polar coordinates grid (corresponding to about one sample per AIA pixel at the limb) with a cadence of one minute.

\subsubsection{Detection statistics\label{sec:detection}}

The power spectral distribution (PSD) is computed independently for each of the $5\times 5$ spatially binned light curves. For each PSD, the significance of a peak of power at a given frequency $\nu$ must be determined with respect to the mean value $\sigma(\nu)$ of the power expected at this frequency from random fluctuations in the absence of a coherent signal \citep[see \S~3 of][]{Auchere2016}. While many coronal time series have power-law-like PSDs \citep{Gruber2011, Auchere2014, Froment2015, Inglis2015, Inglis2016, Ireland2015, Threlfall2017}, finding a generic model of noise able to accurately reproduce all the observed spectral shapes has proved to be challenging \citep{Threlfall2017}. Since periodic signals affect isolated frequency bins, this problem is circumvented in our code by estimating the expected power at each frequency from its average over the 18 neighboring bins. The {\it global}\footnote{{\it Global} confidence levels take into account the total number of degrees of freedom in the spectra, as opposed to {\it local} confidence levels that apply to individual frequencies and/or dates \citep{Auchere2016}.\label{foot:proba}} probability for at least one peak of power greater than $m\sigma(\nu)$ to occur by chance among the $N/2$ frequency bins is then given by
\begin{equation}
P_g(m)=1-\left(1-\text{e}^{-m}\right)^{N/2},
\label{eq:global_proba}
\end{equation}
\noindent

\begin{figure}[t!]
\includegraphics[width=\columnwidth]{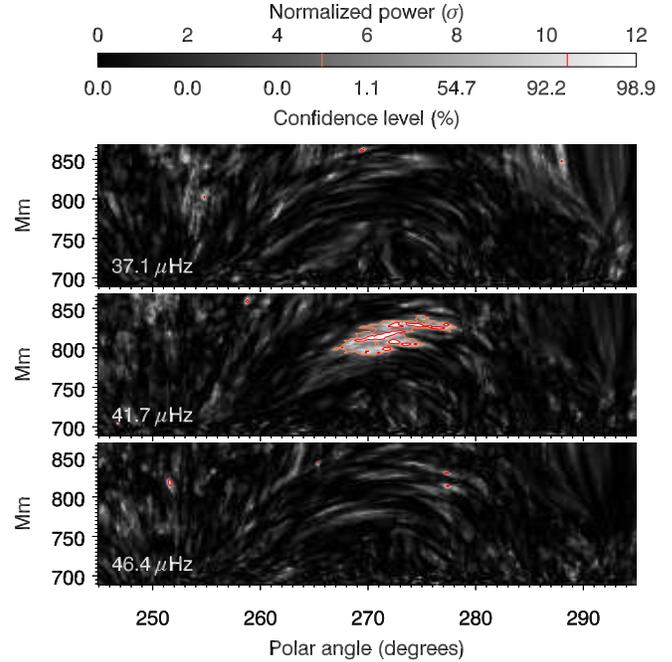}
\caption{Maps of Fourier power at 17.1~nm normalized to its expected value $\sigma(\nu)$ in the absence of a coherent signal, in three contiguous frequency bands. Values above the 95\% confidence level ($10.5 \sigma$, red contours) only have a 5\% chance to be random fluctuations in the corresponding PSDs. They group within the 41.7~\muhz\ frequency band (6.7~hr) in several arc-shaped regions which, along with the surrounding area ($5\sigma$, orange contour), match the top of trans-equatorial loops (Figure~\ref{fig:fov}).\label{fig:power_map}}
\end{figure}

$N=3600$ being the number of data points of the time series \citep{Scargle1982, Gabriel2002, Auchere2016}. Figure~\ref{fig:power_map} gives the 17.1~nm Fourier power normalized to $\sigma(\nu)$ in the ROI (upper color scale) in three contiguous frequency bands (37.1, 41.7, and 46.3 \muhz). The lower color scale gives the {\it global} level of confidence $1-P_g(m)$ that the normalized power $m$ at each point is due to a coherent signal in the corresponding light curve. Due to the large number of frequency bins, it remains below 1\% for powers up to almost $6\sigma$, then quickly rises to reach 95\% at $10.5\sigma$ (red contours) and 99\% at $12.1\sigma$. A group of arc-shaped regions surpassing the gobal 95\% confidence level in the 41.7 \muhz\ band (6.7~hr) outline the top of the trans-equatorial loops of Figure~\ref{fig:fov}. This group is part of a larger area, sharply defined both spatially and in frequency, in which the power is everywhere greater than $5\sigma$ (orange contour), as for the three on-disk cases of \cite{Froment2015}. While the probability to have at least one spectral bin above $5\sigma$ in a PSD is close to one, the probability to have this many contiguous spatial bins above that level within a frequency band is practically equal to zero.

\begin{figure*}[t!]
\includegraphics[width=\textwidth]{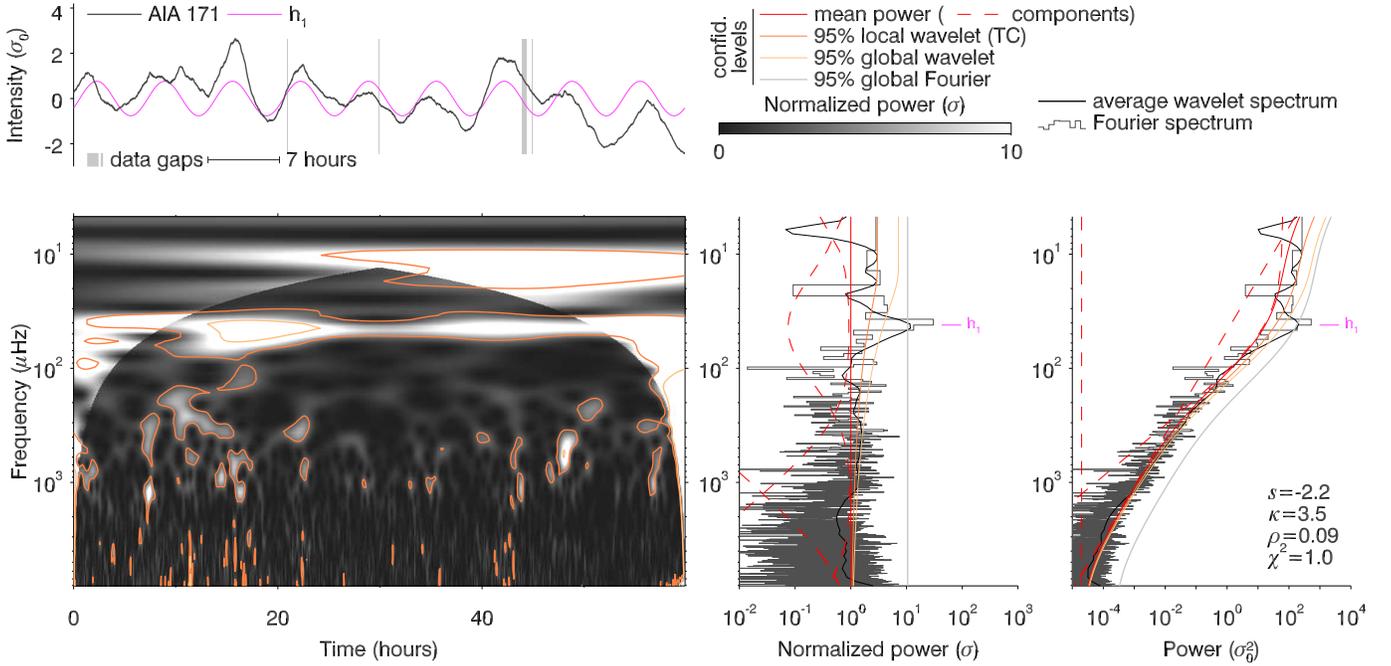}
\caption{The time series averaged over the region of detection (orange contour of Figures~\ref{fig:fov} and~\ref{fig:power_map}) is shown in the top left panel. Its Fourier and time-averaged wavelet power spectra (rightmost panel, gray histograms and black curves) exhibit a broad hump superimposed on a power law leveling off at high frequencies. The $30.2\sigma$ peak of Fourier power labeled h1 at 41.7~\muhz\ stands out in the whitened spectra (middle panel) and has a $1.4\times 10^{-10}$ probability of random occurrence. The corresponding Fourier component is overplotted on the time series in magenta. The whitened wavelet spectrum (left panel) shows a matching strip of significant power lasting for the whole sequence. 
Power within the cone of influence of the Morlet wavelet is shown in lighter shades of gray.\label{fig:fourier_wavelet}}
\end{figure*}

The light curves are then averaged over this region in order to maximize the ratio between the oscillatory signal and the noise. The top left panel of Figure~\ref{fig:fourier_wavelet} shows the resulting 17.1~nm time series normalized to its standard deviation $\sigma_0$. Fourier and Morlet wavelet analysis, performed exactly as described in \cite{Auchere2016, Auchere2016b}, confirm the significance of the detection. The histogram-style curve of the right panel is the Fourier power spectrum of the Hann-apodized time series. The solid red curve is the least-squares fit of this spectrum with the model of power $\sigma(\nu)$ introduced by \citet[][and \S~\ref{sec:pulses} below]{Auchere2016, Auchere2016b}. The peak of Fourier power at 41.7~\muhz\ (6.7~hr) labeled h1 exceeds the 95\% {\it global} confidence level (gray curve) and reaches $30.2\sigma$, which corresponds to a random occurrence probability $P_g(m)=1.4\times 10^{-10}$. The same information is displayed in the middle panel after whitening of the spectrum, i.e. normalization to $\sigma(\nu)$.

The bottom left panel shows the whitened wavelet spectrum of the zero-padded time series. The power at 41.7~\muhz\ exceeds the 95\% local confidence level (orange contours) for the entire duration of the sequence, with a maximum above the 95\% global confidence level (yellow contours) 18~hr after the beginning. Such a long-lived structure has a $4.7\times 10^{-9}$ random occurrence probability. This produces a $11.7\sigma$ peak in the time-averaged wavelet spectrum (black curves in the middle and right panels) that lies above the 95\% global confidence levels (yellow curves), with an associated random occurrence probability of $2.8\times 10^{-10}$.

All confidence levels thus indicate beyond reasonable doubt that the detected periodic signal is real.

\subsubsection{A periodic train of pulses\label{sec:pulses}}

The noise model used to fit the Fourier power spectrum of Figure~\ref{fig:fourier_wavelet} is the sum of three components (dashed red curves in the right panel of Figure~\ref{fig:fourier_wavelet}):
\begin{align}
\sigma(\nu)=A\nu^s+B\text{K}_{\rho,\kappa}(\nu)+C,\nonumber\\
\textrm{with}\ \text{K}_{\rho,\kappa}(\nu)=\left(1+\frac{\nu^2}{\kappa \rho^2}\right)^{-\frac{\kappa+1}{2}}.
\label{eq:kappa_model}
\end{align}
\noindent
The first term is a power-law modeling the background power caused by stochastic processes as observed in the PSDs of most coronal time series. The third one is a constant representing the high frequency white photon noise. The second one is a kappa function centered on zero-frequency introduced by \citet{Auchere2016b} to model the continuum typical of the PSDs of periodic trains of pulses of random amplitudes. This type of signal is periodic in the sense that identical pulses recur at regular intervals, but since their amplitudes are random, the signal is not truly periodic. The randomness of the amplitudes results in the presence of a continuum equal to the PSD of an individual pulse \citep{Xiong2000, Auchere2016b} superimposed on the discrete harmonics expected for a truly periodic signal (with the specific case of the sine curve for which only the fundamental is present).

With a reduced $\chi^2$ of 1.1, the three-component model is a good fit to the Fourier power spectrum of Figure~\ref{fig:fourier_wavelet}. The kappa function hump that dominates the spectrum between 7 and 400~\muhz\ and the strong fundamental peak at 41.7~\muhz\ are signatures characteristic of a train of pulses of random amplitudes. As shown by \citet{Auchere2016b}, higher order harmonics can stand above the noise only if the pulses are pointy enough, i.e. if the kappa function is not too steep.

Simulations of TNE tend to produce nearly identical cycles because they usually impose static boundary conditions \citep[e.g.][]{Karpen2001, Karpen2005, Muller2003, Muller2004, Xia2011, Mikic2013, Fang2015, Froment2017, Froment2018}. In reality however, the heating and the geometry of a loop in TNE are likely to evolve significantly on time-scales shorter than the period. It is thus to be expected that instead of being truly periodic, the corresponding intensity time series present a succession of pulses at least partially decorrelated from one another\footnote{The pulses are susceptible to vary in period, amplitude, and shape. In the above and in \citet{Auchere2016b} we considered only amplitude variations, but the spectral signatures of the other types of random deviations from true periodicity can also be computed \citep{Kaufman1955, Beutler1968}.}. Therefore, as was the case for the two events studied by \citet{Auchere2016b}, without any physical analysis the detailed properties of the PSD already support the TNE scenario.

\subsubsection{The million degree haze\label{sec:haze}}

\begin{figure}[t!]
\includegraphics[width=\columnwidth]{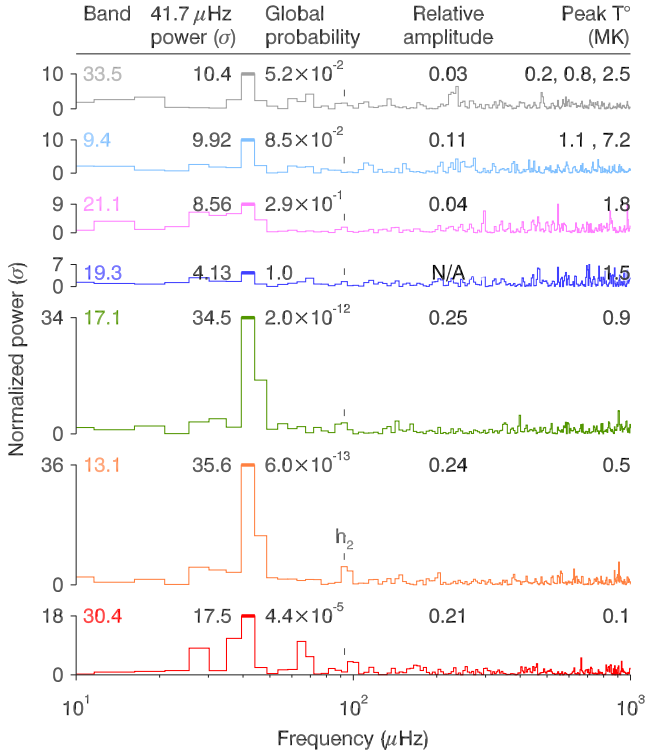}
\caption{Whitened Fourier PSDs of the seven average AIA extreme-ultraviolet light curves corresponding to the orange contour of Figure~\ref{fig:power_map}. The peak of power at 41.7~\muhz\ detected at 17.1~nm (Figures~\ref{fig:power_map} and~\ref{fig:fourier_wavelet}) is absent at 19.3~nm but becomes increasingly significant for the increasingly hotter (21.1, 9.4, and 33.5~nm) and cooler (17.1 and 13.1~nm) coronal bands (see the temperature responses of Figure~\ref{fig:aia_response}), to reach a confidence level of 94.80\% at 33.5~nm and 99.99\% at 13.1~nm. This is explained by the masking of the signal of the pulsating loops bundle around one million degrees by the bulk of the line of sight. The amplitudes are those of the dominant Fourier component relative to the average of the signal. The value at 19.3~nm is meaningless given the global probability of random occurrence.\label{fig:rainbow_psdt}}
\end{figure}

As shown by the whitened PSDs of Figure~\ref{fig:rainbow_psdt}, the peak of power at 41.7~\muhz\ initially detected at 17.1~nm is in fact also present at global significance levels greater than 91\% in four of the six AIA coronal passbands (9.4, 13.1, 17.1, and 33.5~nm)\footnote{Note the possible presence of the second-order harmonic (labeled h$_2$) in the 13.1~nm PSD at 92.9~\muhz. This is 11\% higher than expected but a frequency shift of this amplitude can be explained by the fact that TNE cycles are never exactly periodic (see \S~\ref{sec:pulses}). At $5.4\sigma$, the global probability that this peak is due to random noise is $\approx 1$. However, since the significance of the fundamental was established beforehand, we can use a priori knowledge to restrict the search to within 11\% from the nominal harmonic frequency, which gives a confidence level of 98\% (Equation~\ref{eq:global_proba} with $N=5$).}. The corresponding light curves (obtained by averaging the time series over the contour of detection) are shown in Figure~\ref{fig:rainbow} with the same color coding. The dashed sections correspond to intervals during which the eruptions described in \S~\ref{sec:rain} partially overlap the detection contour. The peak is absent at 19.3~nm and weak at 21.1~nm, the global probabilities of random occurrence being equal to 1.0 and 0.29 respectively. With a maximum response at 1.5~MK (Figure~\ref{fig:aia_response}), the 19.3~nm passband is the most sensitive to the plasma forming the bulk of the background and foreground emission. Indeed, observed off-disk DEMs of active regions are typically centered on 1.5~MK and have a full width at half maximum $\Delta\mathrm{log}\ T_e\approx 0.2$ \citep[e.g.][]{Mason1999, Parenti2000, Parenti2003, Parenti2017, Landi2008, ODwyer2011}. These properties are schematized in Figure~\ref{fig:aia_response} by the shaded Gaussian. 
Therefore, in a given passband, the background signal being proportional to the integral of the product of this DEM by the response function, its contribution relative to that of the pulsating loops is all the greater as the peak of the response -- which shapes the pulses as the temperature periodically drops --  is close to that of the DEM. While this indeed diminishes the relative amplitude of the periodic signal (see the values given in Figure~\ref{fig:rainbow_psdt} and the light curves of Figure~\ref{fig:rainbow}), one should note that if the background emission was constant with time, it would only modify the zero-frequency component of the Fourier PSD. However, since the background emission results from the superimposition of many structures evolving independently and governed by stochastic processes \citep[][]{Ireland2015, Aschwanden2016}, its variance and thus its Fourier power are proportional to its intensity, which ultimately explains why the diminution of the power of the periodic signal normalized to that of the background is maximum in the 19.3~nm passband, as seen in Figure~\ref{fig:rainbow_psdt}.

\subsection{Cooling loops\label{sec:cooling}}

\begin{figure*}[h!]
\includegraphics[width=\textwidth]{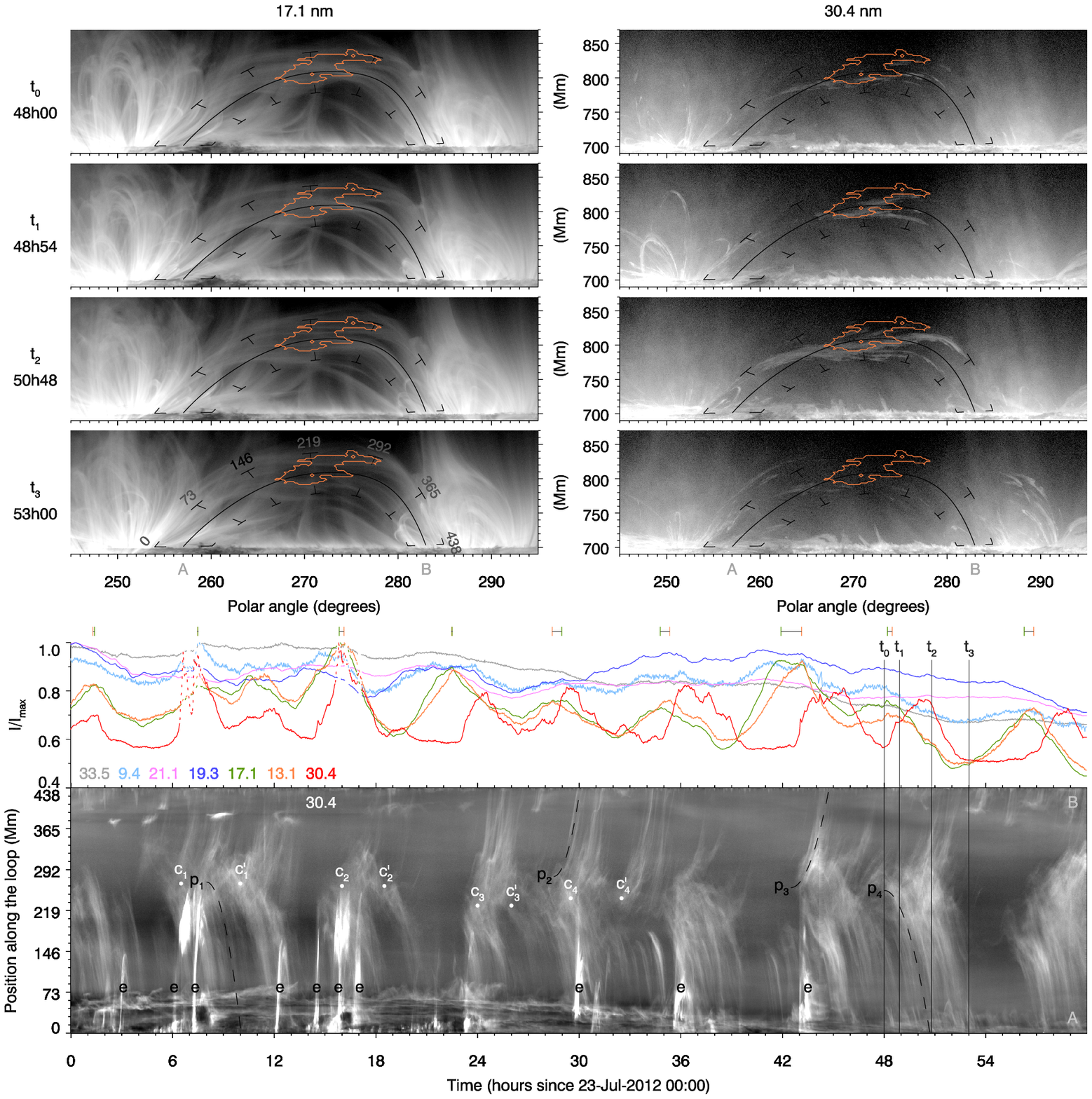}
\caption{Summary of the ``rain bow'' event. The top four lines illustrate the evolution of the intensity at 17.1 (left) and 30.4~nm (right) within the FOV of Figure~\ref{fig:fov} at four successive times ($t_0$ to $t_3$)  during one of the monsoon cycles. The light curves averaged over the region of excess 17.1~nm Fourier power (orange contours) in the seven AIA passbands are plotted below. On average, the local maxima of the hotter 17.1~nm (0.9~MK) light curve (orange ticks) occur before those of the cooler 13.1~nm (0.5~MK) light curve (green ticks), indicating periodic cooling of the plasma. The bottom panel shows the 30.4~nm intensity along the bundle of loops (averaged over the width of the black dashes-delimited bands in the top frames) as a function of time. For every TNE cycle, the temperature drops from several MK (e.g. $t_0$) down to at least 0.09~MK ($t_1$) to form showers of 30.4~nm coronal rain ($t_2$) that drain the loop (t$_3$) and the temperature rises again. See \S\ref{sec:rain} for details. An animated version of this figure is available online.\label{fig:rainbow}}
\end{figure*}

The 41.7 \muhz\ pulsations are most prominent in the 13.1 and 17.1~nm passband (Figures~\ref{fig:rainbow_psdt} and ~\ref{fig:rainbow}). The two light curves are very similar but shifted slightly with respect to each other, 13.1~nm generally peaking after 17.1~nm. The cross-correlation between the two curves has a maximum of 0.96 for a temporal offset of 4 minutes (orange curve of Figure~\ref{fig:rainbow_lags}). A manual estimate of the position of the local maxima in the two passbands (orange and green ticks) gives larger values, with an average delay of 14 minutes.

\begin{figure}[t!]
\includegraphics[width=\columnwidth]{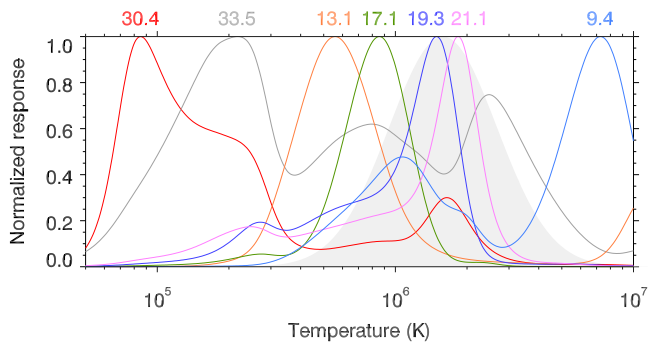}
\caption{Responses of the seven extreme-ultraviolet passbands of AIA to an isothermal plasma as a function of the electron temperature $T_e$, normalized to their maxima. Computed with CHIANTI 8.0.2 \citep{Dere1997, DelZanna2015}, for an electron number density of $10^{15}\ \mathrm{m}^{-3}$. The shaded area represents a typical DEM for the off-disk corona.\label{fig:aia_response}}
\end{figure}

This time lag is a first indication that the plasma of the pulsating bundle of loops is predominantly  seen when cooling. Indeed, when the plasma temperature evolves in response to a heat input, either impulsive \citep{Viall2011, Viall2013} or constant with TNE-prone conditions\footnote{Constant heating without sufficient stratification produces loops in hydrostatic equilibrium whose temperature is thus constant.} \citep{Mikic2013, Froment2017}, the heating phases occur at lower densities than the cooling phases. The temporal evolution of the intensity (which is proportional to the square of the electron density) is thus dominated by the cooling phases. As the temperature decreases, the signal in the AIA passbands describes their temperature responses (Figure~\ref{fig:aia_response}) and thus peaks at 17.1~nm (maximum response at 0.9~MK) before it does at 13.1~nm (maximum at 0.5~MK). Independently of the present periodic behavior, this temporal shift between light curves indicating cooling appears to be a property common to many coronal loops \citep[e.g.][]{Schrijver2001, Ugarte-Urra2006, Ugarte-Urra2009, Warren2007, Landi2009, Kamio2011, Viall2012, Viall2017}.

The periodic signal being visible in five of the six AIA coronal passbands, the cooling of the plasma is traceable through a wide range of temperatures. The time lags corresponding to the maximum cross-correlation between the 17.1~nm light curve and the five other coronal light curves of Figure~\ref{fig:rainbow} are given in Figure~\ref{fig:rainbow_lags}.  
Given its peak response at 1.5~MK, the 19.3~nm passband is oddly found in first position. However, at 19.3~nm, the signal from the pulsating loops is negligible compared to the background and foreground emissions while it represents at least 25\% of the line of sight at 17.1~nm. This is a lower limit based on the relative amplitude of the Fourier component (Figures~\ref{fig:rainbow_psdt} and \ref{fig:rainbow}) and assuming that the loops 17.1~nm emission goes down to zero between pulses. Correlating the 17.1 and 19.3~nm light curves thus amounts to comparing signals coming from separate regions of the corona, hence the low maximum of 0.35. Therefore, for the same reason why the Fourier power is not significant in this passband, the 19.3~nm time lag is not pertinent here, in contrast with the studies performed on disk where the LOS confusion is less \citep{Viall2012, Viall2017, Froment2015}. More generally, the varying contribution of the pulsating loops  to the total LOS intensity is susceptible to affect the interpretation of the cross-correlation time lags between any two light curves \citep[for a detailed discussion, see][]{winebarger2016}.

In order to isolate the contribution of the pulsating loops, \cite{Froment2015} computed instead the phase shift of the main Fourier component for pairs of passbands. The time lags obtained using this method are listed in Figure~\ref{fig:rainbow_lags} above those obtained by cross-correlation. The values are similar but the 33.5~nm passband is now found to peak first, in accordance with its usual hot passband behavior (see below). The method can, however, not completely suppress the influence of the background emission since the phase of the 41.7~\muhz\ Fourier component necessarily contains the contribution from the background at this frequency.

The values of the time lags, and the ordering of the passbands, depend upon the details of the density and temperature evolution, as well as of the shapes of the temperature responses. The 9.4 and 13.1~nm passbands both have two separate response peaks and the 33.5~nm passband has a very broad response with three local maxima (Figure~\ref{fig:aia_response}). These passbands thus do not correspond to a unique temperature or temperature range, which complicates the interpretation of their time lags. In published analyses \citep{Viall2012, Viall2015, Viall2017, Froment2015}, the 13.1~nm passband always has time lags consistent with its main response peak at 0.5~MK, its secondary maximum at 11~MK (off the scale of Figure~\ref{fig:aia_response}) being too weak and/or too hot to contribute significantly to the signal in the nonflaring corona. On the contrary, the 9.4~nm passband is either found to peak first in active region cores (corresponding to its absolute maximum at 7.2~MK) or in the middle of the sequence elsewhere (corresponding to its secondary maximum at 1.1~MK). Despite its maximum response at 0.2~MK\footnote{Caused by far off band \ion{O}{3} to \ion{O}{5} lines that are accounted for only since the 2013 February version 4 update of the AIA response curves above 42.5~nm, based on measurements by \citet{Soufli2012}.\label{foot:reponse}}
, the 33.5~nm passband is usually in the second or first position (when 9.4~nm is not), which corresponds to its secondary maximum at 2.5~MK, because the plasma generally does not cool down much below 0.8~MK \citep{Viall2017}. If it does, however, the 33.5~nm time lags are to be interpreted with caution.

Considering the expected range of variations, the time lags listed in Figure~\ref{fig:rainbow_lags} are consistent with the prevalent cooling pattern revealed by the previous AIA time lag studies, whether on \citep[e.g.][]{Viall2012, Viall2017, Froment2015} or off-disk \citep{Viall2015}. Combined with the periodicity of the signal and its spectral signature as a train of pulses, the present event thus has the same overall characteristics as the ones that \citet{Froment2015, Froment2017} demonstrated as being due to TNE. A notable difference, however, is that while in these latter cases the 13.1 and 17.1~nm light curves had a zero time lag, 13.1~nm is here delayed by 4 to 15 minutes, which indicates that the plasma cools below the maximum response of the 17.1~nm passband at 0.9~MK. 

\subsection{Periodic coronal rain\label{sec:rain}}

\begin{figure}[t!]
\includegraphics[width=\columnwidth]{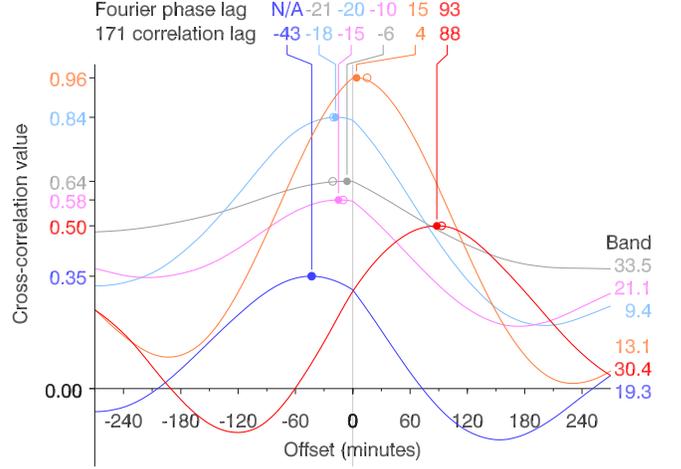}
\caption{Cross-correlation with the 17.1~nm light curve as a function of temporal offset for each of the six other light curves of Figure~\ref{fig:rainbow} (same color coding). The time lags corresponding to the maxima of correlation (left scale) are marked with dots and are listed in ascending order. Above, the phase of the Fourier component at  41.7~\muhz\ (Figure~\ref{fig:rainbow_psdt}) is given for each passband with respect to that of the 17.1~nm passband (open dots). The phase at 19.3~nm is meaningless given the global probability of random ocurrence.\label{fig:rainbow_lags}}
\end{figure}

The 30.4~nm light curve plotted in red in Figure~\ref{fig:rainbow} presents all the properties described above for the coronal passbands. Its Fourier PSD (Figure~\ref{fig:rainbow_psdt}) shows a clear peak at 41.7~\muhz\ and its time delay with respect to 17.1~nm (Figure~\ref{fig:rainbow_lags}) is the greatest. This shows, given the maximum response of the 30.4~nm passband at 0.09~MK (Figure~\ref{fig:aia_response}), that the plasma cools to lower TR temperatures\footnote{This implies that the 0.2~MK peak of the 33.5~nm passband must form a second bump at the end of each pulse in the light curve. However, the relative amplitude of the 33.5~nm pulses is too small for their detailed shape to be analyzed.}. Starting at 23~hr and for all the subsequent cycles, the intensity at 30.4~nm rises immediately after it reaches its maximum at 13.1~nm (0.56~MK). This is consistent with the presence of a broad shoulder\textsuperscript{\ref{foot:reponse}} extending up to 0.25~MK in the temperature response of the 30.4~nm passband.

As can be seen in the animated version of Figure~\ref{fig:rainbow} available online, the 30.4~nm pulses are due to the periodic apparition of cold condensations intersecting the orange detection contour. The top panel of the movie shows a superimposition of simultaneous 30.4, 17.1, and 9.4~nm images in, respectively, the red, green and blue channels of each frame. For easier visual association of passbands with colors, the dimension of the colorspace was reduced by increasing the saturation. With the same color coding, the bottom panel shows the 30.4 (red), 17.1 (green), and 9.4~nm (blue) light curves averaged over the orange contour, the running vertical bar indicating the current time given in the top left corner of each frame. The red 30.4~nm condensations form around the apex of the dominantly green system of loops after the peak of the 17.1~nm intensity.

\subsubsection{Recurring eruptions\label{sec:eruptions}}

Aside from the periodic pulsations and rain in the trans-equatorial bundle of loops, a striking feature of this movie is the recurring confined eruptions originating from the prominence complex superimposed on the left footpoint region. Starting 7~hr after the beginning of the sequence, they occur in phase with six consecutive 30.4~nm pulses. This naturally raises the question of a possible causal relationship between the two phenomena. One could imagine that the eruptions somehow trigger the condensations. But while five of the eruptions occur in the rising phase of the 30.4~nm intensity, the sixth one happens at 26.5~hr during the loop draining of the corresponding cycle, thus after the beginning of the condensations. Furthermore, no eruption takes place during the last two pulses, which rules them out from being the cause. Conversely, one could imagine that the condensations somehow trigger the eruptions. But in this case the latter should occur in the declining phases of the 30.4~nm pulses, and the last two cycles exclude causality also in this direction. It remains the solution that the periodic eruptions are, like the periodic rain, a response of their guiding loops to a quasi-constant and stratified heating, the similar periods being possibly a coincidence and the spatial association only apparent because of foreshortening.

\subsubsection{Multiple bundles\label{sec:rain_los}}

The top four lines of Figure~\ref{fig:rainbow} show the 17.1 (left) and 30.4~nm (right) images at four times, labeled $t_0$ to $t_3$, chosen during the penultimate cycle of the sequence, which is free of eruptions. At $t_0$, the 30.4~nm intensity in the detection contour is minimum, while that at 17.1~nm is near its maximum. The bundle of loops is thus at around 1~MK and cooling. After 54~minutes, at $t_1$, the temperature has decreased past that of the maximum response of the 13.1~nm passband and even down to at least 0.25~MK since the condensations start to be visible at 30.4~nm at the apex of the bundle of loops. At $t_2$ (114~minutes after $t_1$), the condensations are fully developed and flow down toward both footpoints. At $t_3$ (132~minutes after $t_2$), the loops have cooled and drained sufficiently for a darkening at the apex to be clearly visible at 17.1~nm compared to the previous frames. At this point, almost all the condensations have fallen and chromospheric evaporation starts to refill the loops with hot plasma.

It is worth noting that some cold material is present at $t_0$, in between two 30.4~nm pulses. More generally, examination of the movie reveals that there is some rain somewhere along the loops most of the time. This is also clearly visible in the bottom panel of Figure~\ref{fig:rainbow} that shows the 30.4~nm intensity as a function of time along the band delimited by black dashes (top frames) and averaged over its width. The fixed and manually defined band encompasses the detection contour and most of the area of the FOV swept by the system of trans-equatorial loops as a result of their continuously changing shape and position, either due to their intrinsic evolution or to the perspective variations resulting from the $35^{\circ}$ rotation of the Sun during the 2.5 days of the sequence. For better visualization, we compensated the intensity gradient along the loop by subtracting a copy of the image cyclically smoothed over one-third of its length. The bright streaks marked with ``e''s near the ``A'' footpoint are the signatures of the recurring eruptions discussed in \S~\ref{sec:eruptions}. The intervals without any trails of rain amount to 8\% of the total duration, from about 21 to 23~hr,  40 to 41~hr and 53 to 55~hr. Given the width of the bundle, some dispersion in the periods and phases of the large number of individual loops involved is to be expected. Indeed, the groups of apex condensations starting at 6, 15, 19, and 28~hr are each composed of two main clouds (marked with white dots and labeled c$_1^\prime$, c$_1$ to c$_2$, c$_4^\prime$) from which separate showers of rain emerge. Combined with the two distinct northern loop legs anchored on either side of the dashed band at 17.1~nm (see, for example, the movie around 22~hr), this suggests that there are in fact two main bundles of loops superimposed on the LOS, each one made of a number of smaller threads channeling the rain to form the fine individual trails. This is corroborated with a global confidence level of 92\% by the presence of a secondary peak of power\footnote{The third peak of power at 27.9~\muhz\ has a global probability of random occurrence of 0.47 and is thus not significant.} at 65.0~\muhz\ in the PSD of the 30.4~nm light curve. The latter thus results from the superimposition of two periodic signals (41.7 and 65.0~\muhz) possibly corresponding to the two bundles of loops. It, however, remains to be explained why the secondary peak is not visible in the 13.1 or 17.1~nm PSDs in which the primary signal is strong.

\subsubsection{It is raining as usual\label{sec:rain_dynamics}}

To a first approximation, most of the trails of rain in Figure~\ref{fig:rainbow} follow segments of parabolas, indicating a constant net acceleration along the loops. As examples, we overlaid four parabolic trajectories with zero initial velocities: p$_1$ and p$_2$ have an acceleration of 5.5~m.s$^{-2}$, while p$_3$ and p$_4$ have an acceleration of 10~m.s$^{-2}$. The initial section of p$_3$ is not visible, which means that the condensation had already started to fall when it became cool enough to be visible in the 30.4~nm passband. The plane of the sky velocities for p$_1$ to p$_4$ when the drops reach the solar limb are, respectively, 73, 56, 44, and 52 km.s$^{-1}$, which is about four times smaller than the 228~km.s$^{-1}$ expected from the conversion of potential to kinetic energy during a frictionless fall from the apex, 110~Mm above the surface. These kinematic properties are typical of those reported in the literature for other rain events \citep{Schrijver2001, DeGroof2005, Antolin2010, Antolin2011, Antolin2012, Vashalomidze2015}. They are also similar to those found in numerical simulations \citep[e.g][]{Antolin2010, Fang2013, Mikic2013, Xia2017}. Small and constant accelerations along steepening loops, and the small resulting velocities, imply the presence of an upward force that is a combination of drag and either plasma \citep{Oliver2014} or magnetic \citep{Antolin2011, Antolin2015, Verwichte2017} pressure gradient forces.

The exact trajectories are more complex than simple parabolas. For example, possibly bending under the mass of the condensations, the loops develop a dip at their apex toward the end of the sequence. This in turn causes the condensations to linger for a while before falling, forming the wiggly trails starting at 55~hr in the bottom panel of Figure~\ref{fig:rainbow}. This behavior is particularly apparent in the movie and is reminiscent of the formation of a prominence by condensation \citep[e.g.][]{Liu2012, Xia2016}, which is what the TNE mechanism was originally proposed for \citep{Antiochos1991}.

As a final note, the bottom panel of Figure~\ref{fig:rainbow} gives a simplified picture of the temporal evolution due to the averaging over the width of the bundle of loops. A detailed analysis of the rain dynamics would allow further comparison with the predictions of the state of the art 3D MHD models \citep{Moschou2015, Xia2017}, like the deformation of the blobs into V-shapes \citep[as already observed by ][]{Antolin2015} or the presence of an initial Rayleigh-Taylor phase before the fall along the field lines. This would require us to track individual drops and to take into account the as yet unknown three-dimensional geometry, but that would be beyond the scope of this paper. 

\section{The monsoon\label{sec:conclusions}}

In conclusion, the ``rain bow'' event simultaneously exhibits all the attributes of two previously independent phenomena: coronal rain and periodic intensity pulsations of coronal loops. Although the two were already understood to be due to TNE, they had never been observed together before. These new observations thus definitely unify coronal rain and coronal loop pulsations as being two manifestations of the same underlying physical process. The studied period covers eight complete cycles of what can be called -- given the periodicity and extending the meteorological metaphor -- the ``coronal monsoon'': the evaporation of chromospheric plasma followed by its condensation higher up in the loops into ``clouds'' that eventually produce ``rain'' falling back onto the chromosphere.

Simulations \citep{Mikic2013, Froment2018} indicate that depending upon the loop geometry and heating properties, the condensations can be either complete, i.e. cooling to chromospheric temperature and visible as coronal rain, or incomplete, i.e. pushed down one leg by siphon flows before they can fully cool. The two cases reported in \citet{Antolin2015} correspond to the condition of complete condensations. Case~1 of ~\citet{Froment2015} possibly corresponds to the condition of the ``dry'' monsoon, but the difficulty to detect coronal rain on-disk precludes a definitive conclusion. While the basic sequence of processes causing the cycles is established, several aspects of the monsoon events are challenging to understand. It is, for example, surprising that the many loops forming a bundle could evolve in phase. \citet{Antolin2012} already reported a collective behavior of H$\alpha$ rain in neighboring strands, which they suggested to be the result of common footpoint heating conditions. But while this is sufficient to explain similar periods, phasing a priori requires an additional coupling mechanism across the field lines that remains to be identified.
 
Generalizing from coronal rain studies, the main interest of the monsoon events (on or off-disk) is their potential to constrain the location and variability of the coronal heating processes \citep[e.g.][]{Antolin2010}. On the one hand, whatever the actual mechanism, TNE implies that the heating must be sufficiently stratified and quasi-constant but on the other hand, based on simulations of constant cross-section, semicircular vertical loops, \citet{Klimchuk2010} have argued that TNE could not be a widespread mechanism. Several authors, however, showed that if realistic geometries and heating distributions are taken into account, modeled loops can be in a state of TNE while conforming with the coronal heating observational constraints \citep{Lionello2013, Lionello2016, Mikic2013, Winebarger2014}. Using a 1D hydrodynamic model, \citet{Froment2018} have performed for several geometries a systematic study of the occurrence of TNE as a function of the heat flux and of the scale height of the heating deposition. Comparison with the properties of the already large sample of monsoon events will potentially allow the determination of the spatio-temporal statistical distribution of the heating.

It is nonetheless clear that the ``rain bow'' and the on-disk events of \citet{Froment2015} probably do not represent the typical evolution of coronal loops. Being based on Fourier analysis, the method used to detect these events has a strong detection bias toward long-lived and very regular cases. Since thousands of periodic pulsations events have been detected by \citet{Auchere2014} and \citet{Froment2016}, one can therefore predict that many more having less and / or less regular cycles probably exist. Indeed, we note in the movie the presence of many rain events at other locations within the FOV. Examples are visible in the $t_1$ and $t_3$ 30.4~nm frames of Figure~\ref{fig:rainbow} in the two active regions that the ``rain bow'' connects. Quasi-constant heating must thus be present at the base of the corresponding loops but no significant Fourier power could be found in these regions, either because of LOS confusion or because their geometry and/or heating conditions change sufficiently rapidly to prevent TNE cycles from being periodic.

\acknowledgements
AIA data are courtesy of NASA/{\it SDO} and the AIA science team. This work used data provided by the MEDOC data and operations centre (CNES / CNRS / Univ. Paris-Sud), http://medoc.ias.u-psud.fr/.
P.A. has received funding from the UK Science and Technology Facilities Council (Consolidated Grant ST/K000950/1) and the European Union Horizon 2020 research and innovation programme (grant agreement No. 647214).
C.F.: this research was supported by the Research Council of Norway through its Centres of Excellence scheme, project number 262622.
R.O. acknowledges the support from grant AYA2014-54485-P (AEI/FEDER, UE).
We acknowledge support from the International Space Science Institute (ISSI), Bern, Switzerland to the International Team  401 ``Observed Multi-Scale Variability of Coronal Loops as a Probe of Coronal Heating.''

\facility{{it SDO} (AIA)}
\software{Interactive Data Language, CHIANTI \citep{Dere1997}, SolarSoft \citep{Freeland2012}}

\bibliographystyle{aasjournal}
\bibliography{bibliography}

\begin{thebibliography}{}
\expandafter\ifx\csname natexlab\endcsname\relax\def\natexlab#1{#1}\fi
\providecommand{\url}[1]{\href{#1}{#1}}

\bibitem[{{Antiochos} \& {Klimchuk}(1991)}]{Antiochos1991}
{Antiochos}, S.~K., \& {Klimchuk}, J.~A. 1991, \apj, 378, 372

\bibitem[{{Antiochos} {et~al.}(1999){Antiochos}, {MacNeice}, {Spicer}, \&
  {Klimchuk}}]{Antiochos1999}
{Antiochos}, S.~K., {MacNeice}, P.~J., {Spicer}, D.~S., \& {Klimchuk}, J.~A.
  1999, \apj, 512, 985

\bibitem[{{Antolin} \& {Rouppe van der Voort}(2012)}]{Antolin2012}
{Antolin}, P., \& {Rouppe van der Voort}, L. 2012, \apj, 745, 152

\bibitem[{{Antolin} {et~al.}(2010){Antolin}, {Shibata}, \&
  {Vissers}}]{Antolin2010}
{Antolin}, P., {Shibata}, K., \& {Vissers}, G. 2010, \apj, 716, 154

\bibitem[{{Antolin} \& {Verwichte}(2011)}]{Antolin2011}
{Antolin}, P., \& {Verwichte}, E. 2011, \apj, 736, 121

\bibitem[{{Antolin} {et~al.}(2015){Antolin}, {Vissers}, {Pereira}, {Rouppe van
  der Voort}, \& {Scullion}}]{Antolin2015}
{Antolin}, P., {Vissers}, G., {Pereira}, T.~M.~D., {Rouppe van der Voort}, L.,
  \& {Scullion}, E. 2015, \apj, 806, 81

\bibitem[{{Antolin} {et~al.}(2012){Antolin}, {Vissers}, \& {Rouppe van der
  Voort}}]{Antolin2012b}
{Antolin}, P., {Vissers}, G., \& {Rouppe van der Voort}, L. 2012, \solphys,
  280, 457

\bibitem[{{Aschwanden} {et~al.}(2016){Aschwanden}, {Crosby}, {Dimitropoulou},
  {Georgoulis}, {Hergarten}, {McAteer}, {Milovanov}, {Mineshige}, {Morales},
  {Nishizuka}, {Pruessner}, {Sanchez}, {Sharma}, {Strugarek}, \&
  {Uritsky}}]{Aschwanden2016}
{Aschwanden}, M.~J., {Crosby}, N.~B., {Dimitropoulou}, M., {et~al.} 2016, \ssr,
  198, 47

\bibitem[{{Auch{\`e}re} {et~al.}(2014){Auch{\`e}re}, {Bocchialini}, {Solomon},
  \& {Tison}}]{Auchere2014}
{Auch{\`e}re}, F., {Bocchialini}, K., {Solomon}, J., \& {Tison}, E. 2014, \aap,
  563, A8

\bibitem[{{Auch{\`e}re} {et~al.}(2016{\natexlab{a}}){Auch{\`e}re}, {Froment},
  {Bocchialini}, {Buchlin}, \& {Solomon}}]{Auchere2016}
{Auch{\`e}re}, F., {Froment}, C., {Bocchialini}, K., {Buchlin}, E., \&
  {Solomon}, J. 2016{\natexlab{a}}, \apj, 825, 110

\bibitem[{{Auch{\`e}re} {et~al.}(2016{\natexlab{b}}){Auch{\`e}re}, {Froment},
  {Bocchialini}, {Buchlin}, \& {Solomon}}]{Auchere2016b}
---. 2016{\natexlab{b}}, \apj, 827, 152

\bibitem[{{Beutler} \& {Leneman}(1968)}]{Beutler1968}
{Beutler}, F.~J., \& {Leneman}, O.~A.~Z. 1968, Information and Control, 12, 236

\bibitem[{{De Groof} {et~al.}(2005){De Groof}, {Bastiaensen}, {M{\"u}ller},
  {Berghmans}, \& {Poedts}}]{DeGroof2005}
{De Groof}, A., {Bastiaensen}, C., {M{\"u}ller}, D.~A.~N., {Berghmans}, D., \&
  {Poedts}, S. 2005, \aap, 443, 319

\bibitem[{{De Groof} {et~al.}(2004){De Groof}, {Berghmans}, {van
  Driel-Gesztelyi}, \& {Poedts}}]{DeGroof2004}
{De Groof}, A., {Berghmans}, D., {van Driel-Gesztelyi}, L., \& {Poedts}, S.
  2004, \aap, 415, 1141

\bibitem[{{Del Zanna} {et~al.}(2015){Del Zanna}, {Dere}, {Young}, {Landi}, \&
  {Mason}}]{DelZanna2015}
{Del Zanna}, G., {Dere}, K.~P., {Young}, P.~R., {Landi}, E., \& {Mason}, H.~E.
  2015, \aap, 582, A56

\bibitem[{{Delaboudini{\`e}re} {et~al.}(1995){Delaboudini{\`e}re}, {Artzner},
  {Brunaud}, {Gabriel}, {Hochedez}, {Millier}, {Song}, {Au}, {Dere}, {Howard},
  {Kreplin}, {Michels}, {Moses}, {Defise}, {Jamar}, {Rochus}, {Chauvineau},
  {Marioge}, {Catura}, {Lemen}, {Shing}, {Stern}, {Gurman}, {Neupert},
  {Maucherat}, {Clette}, {Cugnon}, \& {van Dessel}}]{Delaboudiniere1995}
{Delaboudini{\`e}re}, J.-P., {Artzner}, G.~E., {Brunaud}, J., {et~al.} 1995,
  \solphys, 162, 291

\bibitem[{{Dere} {et~al.}(1997){Dere}, {Landi}, {Mason}, {Monsignori Fossi}, \&
  {Young}}]{Dere1997}
{Dere}, K.~P., {Landi}, E., {Mason}, H.~E., {Monsignori Fossi}, B.~C., \&
  {Young}, P.~R. 1997, \aaps, 125, 149

\bibitem[{{Domingo} {et~al.}(1995){Domingo}, {Fleck}, \&
  {Poland}}]{Domingo1995}
{Domingo}, V., {Fleck}, B., \& {Poland}, A.~I. 1995, \solphys, 162, 1

\bibitem[{{Fang} {et~al.}(2013){Fang}, {Xia}, \& {Keppens}}]{Fang2013}
{Fang}, X., {Xia}, C., \& {Keppens}, R. 2013, \apjl, 771, L29

\bibitem[{{Fang} {et~al.}(2015){Fang}, {Xia}, {Keppens}, \& {Van
  Doorsselaere}}]{Fang2015}
{Fang}, X., {Xia}, C., {Keppens}, R., \& {Van Doorsselaere}, T. 2015, \apj,
  807, 142

\bibitem[{{Field}(1965)}]{Field1965}
{Field}, G.~B. 1965, \apj, 142, 531

\bibitem[{{Freeland} \& {Handy}(2012)}]{Freeland2012}
{Freeland}, S.~L., \& {Handy}, B.~N. 2012, {SolarSoft: Programming and data
  analysis environment for solar physics}, Astrophysics Source Code Library, ,
  , ascl:1208.013

\bibitem[{{Froment}(2016)}]{Froment2016}
{Froment}, C. 2016, PhD thesis, Universit{\'e} Paris-Saclay, Universit{\'e}
  Paris-Sud, Institut d'Astrophysique Spatiale, Orsay, France

\bibitem[{{Froment} {et~al.}(2017){Froment}, {Auch{\`e}re}, {Aulanier},
  {Miki{\'c}}, {Bocchialini}, {Buchlin}, \& {Solomon}}]{Froment2017}
{Froment}, C., {Auch{\`e}re}, F., {Aulanier}, G., {et~al.} 2017, \apj, 835, 272

\bibitem[{{Froment} {et~al.}(2018){Froment}, {Auch{\`e}re}, {Aulanier},
  {Miki{\'c}}, {Bocchialini}, {Buchlin}, \& {Solomon}}]{Froment2018}
---. 2018, \apj, submitted

\bibitem[{{Froment} {et~al.}(2015){Froment}, {Auch{\`e}re}, {Bocchialini},
  {Buchlin}, {Guennou}, \& {Solomon}}]{Froment2015}
{Froment}, C., {Auch{\`e}re}, F., {Bocchialini}, K., {et~al.} 2015, \apj, 807,
  158

\bibitem[{{Gabriel} {et~al.}(2002){Gabriel}, {Baudin}, {Boumier},
  {Garc{\'{\i}}a}, {Turck-Chi{\`e}ze}, {Appourchaux}, {Bertello}, {Berthomieu},
  {Charra}, {Gough}, {Pall{\'e}}, {Provost}, {Renaud}, {Robillot}, {Roca
  Cort{\'e}s}, {Thiery}, \& {Ulrich}}]{Gabriel2002}
{Gabriel}, A.~H., {Baudin}, F., {Boumier}, P., {et~al.} 2002, \aap, 390, 1119

\bibitem[{{Gruber} {et~al.}(2011){Gruber}, {Lachowicz}, {Bissaldi}, {Briggs},
  {Connaughton}, {Greiner}, {van der Horst}, {Kanbach}, {Rau}, {Bhat}, {Diehl},
  {von Kienlin}, {Kippen}, {Meegan}, {Paciesas}, {Preece}, \&
  {Wilson-Hodge}}]{Gruber2011}
{Gruber}, D., {Lachowicz}, P., {Bissaldi}, E., {et~al.} 2011, \aap, 533, A61

\bibitem[{{Gudiksen} \& {Nordlund}(2005)}]{Gudiksen2005b}
{Gudiksen}, B.~V., \& {Nordlund}, {\AA}. 2005, \apj, 618, 1031

\bibitem[{{Guennou} {et~al.}(2013){Guennou}, {Auch{\`e}re}, {Klimchuk},
  {Bocchialini}, \& {Parenti}}]{Guennou2013}
{Guennou}, C., {Auch{\`e}re}, F., {Klimchuk}, J.~A., {Bocchialini}, K., \&
  {Parenti}, S. 2013, \apj, 774, 31

\bibitem[{{Guennou} {et~al.}(2012{\natexlab{a}}){Guennou}, {Auch{\`e}re},
  {Soubri{\'e}}, {Bocchialini}, {Parenti}, \& {Barbey}}]{Guennou2012a}
{Guennou}, C., {Auch{\`e}re}, F., {Soubri{\'e}}, E., {et~al.}
  2012{\natexlab{a}}, \apjs, 203, 25

\bibitem[{{Guennou} {et~al.}(2012{\natexlab{b}}){Guennou}, {Auch{\`e}re},
  {Soubri{\'e}}, {Bocchialini}, {Parenti}, \& {Barbey}}]{Guennou2012b}
---. 2012{\natexlab{b}}, \apjs, 203, 26

\bibitem[{{Inglis} {et~al.}(2016){Inglis}, {Ireland}, {Dennis}, {Hayes}, \&
  {Gallagher}}]{Inglis2016}
{Inglis}, A.~R., {Ireland}, J., {Dennis}, B.~R., {Hayes}, L., \& {Gallagher},
  P. 2016, \apj, 833, 284

\bibitem[{{Inglis} {et~al.}(2015){Inglis}, {Ireland}, \&
  {Dominique}}]{Inglis2015}
{Inglis}, A.~R., {Ireland}, J., \& {Dominique}, M. 2015, \apj, 798, 108

\bibitem[{{Ireland} {et~al.}(2015){Ireland}, {McAteer}, \&
  {Inglis}}]{Ireland2015}
{Ireland}, J., {McAteer}, R.~T.~J., \& {Inglis}, A.~R. 2015, \apj, 798, 1

\bibitem[{{Kamio} {et~al.}(2011){Kamio}, {Peter}, {Curdt}, \&
  {Solanki}}]{Kamio2011}
{Kamio}, S., {Peter}, H., {Curdt}, W., \& {Solanki}, S.~K. 2011, \aap, 532, A96

\bibitem[{{Karpen} \& {Antiochos}(2008)}]{Karpen2008}
{Karpen}, J.~T., \& {Antiochos}, S.~K. 2008, \apj, 676, 658

\bibitem[{{Karpen} {et~al.}(2001){Karpen}, {Antiochos}, {Hohensee}, {Klimchuk},
  \& {MacNeice}}]{Karpen2001}
{Karpen}, J.~T., {Antiochos}, S.~K., {Hohensee}, M., {Klimchuk}, J.~A., \&
  {MacNeice}, P.~J. 2001, \apjl, 553, L85

\bibitem[{{Karpen} {et~al.}(2005){Karpen}, {Tanner}, {Antiochos}, \&
  {DeVore}}]{Karpen2005}
{Karpen}, J.~T., {Tanner}, S.~E.~M., {Antiochos}, S.~K., \& {DeVore}, C.~R.
  2005, \apj, 635, 1319

\bibitem[{{Kaufman} \& {King}(1955)}]{Kaufman1955}
{Kaufman}, H., \& {King}, E.~H. 1955, IRE Transactions on Information Theory,
  1, 40

\bibitem[{{Kawaguchi}(1970)}]{Kawaguchi1970}
{Kawaguchi}, I. 1970, \pasj, 22, 405

\bibitem[{{Klimchuk} {et~al.}(2010){Klimchuk}, {Karpen}, \&
  {Antiochos}}]{Klimchuk2010}
{Klimchuk}, J.~A., {Karpen}, J.~T., \& {Antiochos}, S.~K. 2010, \apj, 714, 1239

\bibitem[{{Kuin} \& {Martens}(1982)}]{Kuin1982}
{Kuin}, N.~P.~M., \& {Martens}, P.~C.~H. 1982, \aap, 108, L1

\bibitem[{{Landi} \& {Feldman}(2008)}]{Landi2008}
{Landi}, E., \& {Feldman}, U. 2008, \apj, 672, 674

\bibitem[{{Landi} {et~al.}(2009){Landi}, {Miralles}, {Curdt}, \&
  {Hara}}]{Landi2009}
{Landi}, E., {Miralles}, M.~P., {Curdt}, W., \& {Hara}, H. 2009, \apj, 695, 221

\bibitem[{{Lemen} {et~al.}(2012){Lemen}, {Title}, {Akin}, {Boerner}, {Chou},
  {Drake}, {Duncan}, {Edwards}, {Friedlaender}, {Heyman}, {Hurlburt}, {Katz},
  {Kushner}, {Levay}, {Lindgren}, {Mathur}, {McFeaters}, {Mitchell}, {Rehse},
  {Schrijver}, {Springer}, {Stern}, {Tarbell}, {Wuelser}, {Wolfson}, {Yanari},
  {Bookbinder}, {Cheimets}, {Caldwell}, {Deluca}, {Gates}, {Golub}, {Park},
  {Podgorski}, {Bush}, {Scherrer}, {Gummin}, {Smith}, {Auker}, {Jerram},
  {Pool}, {Soufli}, {Windt}, {Beardsley}, {Clapp}, {Lang}, \&
  {Waltham}}]{Lemen2012}
{Lemen}, J.~R., {Title}, A.~M., {Akin}, D.~J., {et~al.} 2012, \solphys, 275, 17

\bibitem[{{Leroy}(1972)}]{Leroy1972}
{Leroy}, J.-L. 1972, \solphys, 25, 413

\bibitem[{{Lionello} {et~al.}(2016){Lionello}, {Alexander}, {Winebarger},
  {Linker}, \& {Miki{\'c}}}]{Lionello2016}
{Lionello}, R., {Alexander}, C.~E., {Winebarger}, A.~R., {Linker}, J.~A., \&
  {Miki{\'c}}, Z. 2016, \apj, 818, 129

\bibitem[{{Lionello} {et~al.}(2013){Lionello}, {Winebarger}, {Mok}, {Linker},
  \& {Miki{\'c}}}]{Lionello2013}
{Lionello}, R., {Winebarger}, A.~R., {Mok}, Y., {Linker}, J.~A., \&
  {Miki{\'c}}, Z. 2013, \apj, 773, 134

\bibitem[{{Liu} {et~al.}(2012){Liu}, {Berger}, \& {Low}}]{Liu2012}
{Liu}, W., {Berger}, T.~E., \& {Low}, B.~C. 2012, \apjl, 745, L21

\bibitem[{{Martens} \& {Kuin}(1983)}]{Martens1983}
{Martens}, P.~C.~H., \& {Kuin}, N.~P.~M. 1983, \aap, 123, 216

\bibitem[{{Mason} {et~al.}(1999){Mason}, {Landi}, {Pike}, \&
  {Young}}]{Mason1999}
{Mason}, H.~E., {Landi}, E., {Pike}, C.~D., \& {Young}, P.~R. 1999, \solphys,
  189, 129

\bibitem[{{Miki{\'c}} {et~al.}(2013){Miki{\'c}}, {Lionello}, {Mok}, {Linker},
  \& {Winebarger}}]{Mikic2013}
{Miki{\'c}}, Z., {Lionello}, R., {Mok}, Y., {Linker}, J.~A., \& {Winebarger},
  A.~R. 2013, \apj, 773, 94

\bibitem[{{Moschou} {et~al.}(2015){Moschou}, {Keppens}, {Xia}, \&
  {Fang}}]{Moschou2015}
{Moschou}, S.~P., {Keppens}, R., {Xia}, C., \& {Fang}, X. 2015, Advances in
  Space Research, 56, 2738

\bibitem[{{M{\"u}ller} {et~al.}(2003){M{\"u}ller}, {Hansteen}, \&
  {Peter}}]{Muller2003}
{M{\"u}ller}, D.~A.~N., {Hansteen}, V.~H., \& {Peter}, H. 2003, \aap, 411, 605

\bibitem[{{M{\"u}ller} {et~al.}(2004){M{\"u}ller}, {Peter}, \&
  {Hansteen}}]{Muller2004}
{M{\"u}ller}, D.~A.~N., {Peter}, H., \& {Hansteen}, V.~H. 2004, \aap, 424, 289

\bibitem[{{O'Dwyer} {et~al.}(2011){O'Dwyer}, {Del Zanna}, {Mason}, {Sterling},
  {Tripathi}, \& {Young}}]{ODwyer2011}
{O'Dwyer}, B., {Del Zanna}, G., {Mason}, H.~E., {et~al.} 2011, \aap, 525, A137

\bibitem[{{Oliver} {et~al.}(2014){Oliver}, {Soler}, {Terradas}, {Zaqarashvili},
  \& {Khodachenko}}]{Oliver2014}
{Oliver}, R., {Soler}, R., {Terradas}, J., {Zaqarashvili}, T.~V., \&
  {Khodachenko}, M.~L. 2014, \apj, 784, 21

\bibitem[{{Parenti} {et~al.}(2000){Parenti}, {Bromage}, {Poletto}, {Noci},
  {Raymond}, \& {Bromage}}]{Parenti2000}
{Parenti}, S., {Bromage}, B.~J.~I., {Poletto}, G., {et~al.} 2000, \aap, 363,
  800

\bibitem[{{Parenti} {et~al.}(2017){Parenti}, {del Zanna}, {Petralia}, {Reale},
  {Teriaca}, {Testa}, \& {Mason}}]{Parenti2017}
{Parenti}, S., {del Zanna}, G., {Petralia}, A., {et~al.} 2017, \apj, 846, 25

\bibitem[{{Parenti} {et~al.}(2003){Parenti}, {Landi}, \&
  {Bromage}}]{Parenti2003}
{Parenti}, S., {Landi}, E., \& {Bromage}, B.~J.~I. 2003, \apj, 590, 519

\bibitem[{{Parker}(1953)}]{Parker1953}
{Parker}, E.~N. 1953, \apj, 117, 431

\bibitem[{{Pesnell} {et~al.}(2012){Pesnell}, {Thompson}, \&
  {Chamberlin}}]{Pesnell2012}
{Pesnell}, W.~D., {Thompson}, B.~J., \& {Chamberlin}, P.~C. 2012, \solphys,
  275, 3

\bibitem[{{Reale}(2014)}]{Reale2014}
{Reale}, F. 2014, Living Reviews in Solar Physics, 11, doi:10.12942/lrsp-2014-4

\bibitem[{{Scargle}(1982)}]{Scargle1982}
{Scargle}, J.~D. 1982, \apj, 263, 835

\bibitem[{{Schrijver}(2001)}]{Schrijver2001}
{Schrijver}, C.~J. 2001, \solphys, 198, 325

\bibitem[{{Scullion} {et~al.}(2016){Scullion}, {Rouppe van der Voort},
  {Antolin}, {Wedemeyer}, {Vissers}, {Kontar}, \& {Gallagher}}]{Scullion2016}
{Scullion}, E., {Rouppe van der Voort}, L., {Antolin}, P., {et~al.} 2016, \apj,
  833, 184

\bibitem[{{Soufli} {et~al.}(2012){Soufli}, {Spiller}, {Windt}, {Robinson},
  {Rodriguez-de Marcos}, {Fernandez-Perea}, {Baker}, {Aquila}, {Dollar},
  {M{\'e}ndez}, {Larruquert}, {Golub}, \& {Boerner}}]{Soufli2012}
{Soufli}, R., {Spiller}, E., {Windt}, D.~L., {et~al.} 2012, in \procspie, Vol.
  8443, Space Telescopes and Instrumentation 2012: Ultraviolet to Gamma Ray,
  84433C

\bibitem[{{Threlfall} {et~al.}(2017){Threlfall}, {De Moortel}, \&
  {Conlon}}]{Threlfall2017}
{Threlfall}, J., {De Moortel}, I., \& {Conlon}, T. 2017, \solphys, 292, 165

\bibitem[{{Ugarte-Urra} {et~al.}(2009){Ugarte-Urra}, {Warren}, \&
  {Brooks}}]{Ugarte-Urra2009}
{Ugarte-Urra}, I., {Warren}, H.~P., \& {Brooks}, D.~H. 2009, \apj, 695, 642

\bibitem[{{Ugarte-Urra} {et~al.}(2006){Ugarte-Urra}, {Winebarger}, \&
  {Warren}}]{Ugarte-Urra2006}
{Ugarte-Urra}, I., {Winebarger}, A.~R., \& {Warren}, H.~P. 2006, \apj, 643,
  1245

\bibitem[{{Vashalomidze} {et~al.}(2015){Vashalomidze}, {Kukhianidze},
  {Zaqarashvili}, {Oliver}, {Shergelashvili}, {Ramishvili}, {Poedts}, \& {De
  Causmaecker}}]{Vashalomidze2015}
{Vashalomidze}, Z., {Kukhianidze}, V., {Zaqarashvili}, T.~V., {et~al.} 2015,
  \aap, 577, A136

\bibitem[{{Verwichte} {et~al.}(2017){Verwichte}, {Antolin}, {Rowlands},
  {Kohutova}, \& {Neukirch}}]{Verwichte2017}
{Verwichte}, E., {Antolin}, P., {Rowlands}, G., {Kohutova}, P., \& {Neukirch},
  T. 2017, \aap, 598, A57

\bibitem[{{Viall} \& {Klimchuk}(2011)}]{Viall2011}
{Viall}, N.~M., \& {Klimchuk}, J.~A. 2011, \apj, 738, 24

\bibitem[{{Viall} \& {Klimchuk}(2012)}]{Viall2012}
---. 2012, \apj, 753, 35

\bibitem[{{Viall} \& {Klimchuk}(2013)}]{Viall2013}
---. 2013, The Astrophysical Journal, 771, 115

\bibitem[{{Viall} \& {Klimchuk}(2015)}]{Viall2015}
---. 2015, \apj, 799, 58

\bibitem[{{Viall} \& {Klimchuk}(2017)}]{Viall2017}
---. 2017, \apj, 842, 108

\bibitem[{{Warren} {et~al.}(2007){Warren}, {Ugarte-Urra}, {Brooks}, {Cirtain},
  {Williams}, \& {Hara}}]{Warren2007}
{Warren}, H.~P., {Ugarte-Urra}, I., {Brooks}, D.~H., {et~al.} 2007, \pasj, 59,
  S675

\bibitem[{{Winebarger} {et~al.}(2016){Winebarger}, {Lionello}, {Downs},
  {Miki{\'c}}, {Linker}, \& {Mok}}]{winebarger2016}
{Winebarger}, A.~R., {Lionello}, R., {Downs}, C., {et~al.} 2016, \apj, 831, 172

\bibitem[{{Winebarger} {et~al.}(2014){Winebarger}, {Lionello}, {Mok}, {Linker},
  \& {Miki{\'c}}}]{Winebarger2014}
{Winebarger}, A.~R., {Lionello}, R., {Mok}, Y., {Linker}, J.~A., \&
  {Miki{\'c}}, Z. 2014, \apj, 795, 138

\bibitem[{{Xia} {et~al.}(2011){Xia}, {Chen}, {Keppens}, \& {van
  Marle}}]{Xia2011}
{Xia}, C., {Chen}, P.~F., {Keppens}, R., \& {van Marle}, A.~J. 2011, \apj, 737,
  27

\bibitem[{{Xia} \& {Keppens}(2016)}]{Xia2016}
{Xia}, C., \& {Keppens}, R. 2016, \apj, 823, 22

\bibitem[{{Xia} {et~al.}(2017){Xia}, {Keppens}, \& {Fang}}]{Xia2017}
{Xia}, C., {Keppens}, R., \& {Fang}, X. 2017, \aap, 603, A42

\bibitem[{{Xiong}(2000)}]{Xiong2000}
{Xiong}, F. 2000, {Digital Modulation Techniques} ({Artech House})

\end{thebibliography}

\end{document}